\documentclass[journal]{IEEEtran}
\usepackage[latin1]{inputenc}
\usepackage{times,amsmath}
\usepackage{amssymb}
\usepackage{steinmetz}
\usepackage{pstool}
\usepackage{subfigure}
\usepackage{multirow}
\usepackage{enumerate}
\usepackage{graphicx}
\usepackage{multirow}
\usepackage{subfig}
\usepackage{mwe}
\usepackage{bigints}
\usepackage{MnSymbol}
\usepackage{stfloats} 
\usepackage[table]{xcolor}
\usepackage[square, comma, sort&compress, numbers]{natbib}
\usepackage{nohyperref}
\usepackage{algorithm,algorithmic}
\usepackage{epstopdf}
\usepackage{mathtools, cuted}

\graphicspath{ {Figures/} }

\begin{document}

\title{ Hand-breathe: Non-Contact Monitoring of Breathing Abnormalities from Hand Palm}

\author{
\IEEEauthorblockN{
Kawish Pervez\IEEEauthorrefmark{1},Waqas Aman\IEEEauthorrefmark{2}, M. Mahboob Ur Rahman\IEEEauthorrefmark{1}, M. Wasim Nawaz\IEEEauthorrefmark{3}, Qammer H. Abbasi\IEEEauthorrefmark{4} }

\IEEEauthorblockA{\IEEEauthorrefmark{1} Electrical engineering department, Information Technology University, Lahore 54000, Pakistan\\ \IEEEauthorrefmark{2} College of Science and Engineering, Hamad Bin Khalifa University (HBKU), Doha, Qatar \\
\IEEEauthorrefmark{3} Department of Computer Engineering,
The University of Lahore, Lahore, 54000, Pakistan
\\ \IEEEauthorrefmark{4}Department of Electronics and Nano Engineering, University of Glasgow, Glasgow, G12 8QQ, UK\\
\IEEEauthorrefmark{1} \{msee20045, mahboob.rahman\}@itu.edu.pk, \IEEEauthorrefmark{2} waman@hbku.edu.qa, \IEEEauthorrefmark{3} muhammad.wasim@dce.uol.edu.pk, \\ \IEEEauthorrefmark{4} Qammer.Abbasi@glasgow.ac.uk }
}

\maketitle

\begin{abstract} 
 
In post-covid19 world, an established measure of caution is to avoid frequent contact with humans, and with surfaces of shared things. At the same time, covid19 has also acted as a catalyst to increase the awareness and concern among the general public to self-assess their overall well-being every now and then. Therefore, the world has witnessed an increased demand for diagnostic tools for contact-less monitoring of body vitals (e.g., infrared-based temperature guns). In this backdrop, radio frequency (RF)-based non-contact methods, e.g., software-defined radios (SDR)-based methods are promising candidates for intelligent remote sensing of human vitals, and could help in containment of contagious viruses like covid19. To this end, this work utilizes the universal software radio peripherals (USRP)-based SDRs along with classical machine learning methods to design a non-contact method to monitor different breathing patterns (including breathing abnormalities). Under our proposed method, a subject rests his/her hand on a table in between the transmit and receive antennas, while an orthogonal frequency division multiplexing (OFDM) signal passes through the hand. Subsequently, the receiver extracts the channel frequency response (basically, fine-grained wireless channel state information), and feeds it to various ML algorithms which eventually classify between different breathing abnormalities. For the proposed method, among all classifiers, linear SVM classifier resulted in a maximum accuracy of 88.1\%. To train the ML classifiers in a supervised manner, data was collected by doing real-time experiments on 4 subjects in a lab environment. For label generation purpose, the breathing of the subjects was classified into three classes: normal, fast, and slow breathing. Furthermore, in addition to our proposed method (where only a hand is exposed to RF signals), we also implemented and tested the state-of-the-art method (where full chest is exposed to RF radiation). The performance comparison of the two methods reveals a trade-off, i.e., the accuracy of our proposed method is slightly inferior than the benchmark method, but our method results in minimal body exposure to RF radiation compared to the benchmark method. Thus, the proposed method is quite amenable for rapid testing of masses for covid19 in particular, and many respiratory disorders in general).  

\end{abstract}

\section{Introduction}
\label{sec:intro}

Breathing (or, respiratory) rate is one of the most important vital signs that needs to be monitored from time to time. In fact, breathing rate becomes relatively more significant compared to some other vitals (e.g., blood pressure) in ascertaining one's overall well-being (e.g., after general anesthesia) \cite{goldhill2005physiologically}. According to a World Health Organization (WHO) study, respiratory abnormalities are one of the leading causes of disability and death worldwide\cite{levine2022global}. The risk factors for abnormal breathing include smoking, air pollution, malnutrition, and others. Breathing abnormality, when occurs, could indicate the following: cardiac arrest, covid19 \cite{subbe2003effect}, \cite{fieselmann1993respiratory}, chronic obstructive pulmonary disease (COPD), asthma, and lungs cancer\cite{smith1993fuel,farida2020quality}. Thus, continuous (or periodic) monitoring of breathing patterns could help raise an early alarm to prompt the medical experts to differentially diagnose the underlying problem. 

A very brief pathophysiology of breathing disorder is as follows. The breathing cycle includes inspiration of $O_2$ to the expiration of $CO_2$. This exchange of gases takes place in the alveolar membrane \cite{tu2012computational,garcia2018computational}. Air from the lungs travels from the larynx and trachea to the nasal, pharyngeal, and oral cavities during exhale. Thus, the quality of our voice is influenced by our exhalation's strength, variety, and speed (depending on the amount of syllables in our speech). There is coordination in the respiratory system with the laryngeal-based subsystems \cite{zhang2016respiratory,gramming1988relationship}. During asymptomatic stages of diseases, the changes might be minor but they eventually affect the vocal coordination and subsystems. Taking covid19 as an example, the respiratory system as well  as the neurological system may be affected by the disease. Additionally, covid19 affects the function of the respiratory process that involves the diaphragm and respiratory tract. As a result, the pattern of exhaling and inhaling from lungs is affected and leads to breathing abnormalities \cite{world2020clinical}. 

The breathing rate and other body vitals have traditionally been measured using either invasive (e.g., capnography, catheters) or contact-based methods (e.g., electrocardiographs, pulse-oximeters, etc.) until very recently. But ever since the outbreak of covid19, the clinical context has changed drastically. Specifically, it is now well-known that the covid19 pathogen/virus could stay on various surfaces (e.g., plastic, metal) for days, and could possibly infect a healthy person upon touching such surfaces. This fact alone has motivated the researchers to devise novel mechanisms to measure the body vitals (e.g., heart rate, blood oxygen saturation level, breathing rate, etc.) in a non-contact (and non-invasive) manner\footnote{Some further disadvantages of contact-based methods are as follows. Continuous use of contact-based sensors (e.g., ECG electrodes, pulse-oximeters, smart watches) could cause skin-related problems and create discomfort as well. Persons with dementia may forget to put on the wearable sensor.}. To this end, a range of technologies have been proposed to design contactless solutions for monitoring of body vitals which could help spread the spread of the contagious covid19 disease. Additionally, such tools have the potential to reduce the dependence of the patients on the visits to the hospitals, and thus, help share the burden of the health care systems \cite{saeed2021wireless}. Finally, non-contact methods could monitor human vitals from a distance, thus, they allow long-term and real-time monitoring of a subject without any potential inconvenience \cite{yatani2012bodyscope, ertin2011autosense}.  	

To date, various promising solutions for non-contact sensing of body vitals have been proposed \cite{abdelnasser2015wigest,wang2014we,wang2016lifs,lien2016soli}, which could be categorized into the following four classes. 

1) Camera-based monitoring systems: Such methods record video of a subject from a distance and estimate vitals by exploiting the periodic change of skin color to measure the heart rate and measure the periodic chest movement to estimate the breathing rate. Despite the promise, such systems have some limitations, e.g., they require good light around the surrounding, can't see through the walls, and have a restricted viewing angle, etc. 

2) Radar-based monitoring systems: These systems use classical radar principles, i.e., range and doppler, to estimate the vitals. Such systems are accurate (especially at mm-wave frequencies), could localize the target (i.e., heart that generates the pulse). But the cost of equipment, and radiation hazards are the limitations of such systems \cite{10.1145/2829988.2787487}. 

3) Wi-Fi-based monitoring systems: This method utilizes the existing widespread infrastructure of WiFi routers in indoor environments. With little modification, a WiFi router could collect the signals reflected-off a person in order to estimate body vitals using the cutting-edge machine learning (ML) and deep learning (DL) methods. This method is cost-effective, and could prove to be ubiquitous, but has some drawbacks too, e.g., lack of flexibility (due to rigid hardware design). 

4) Software-defined radio-based monitoring systems: Such systems utilize a pair of software-defined radios (SDR) in order to do contact-less monitoring of body vitals, and offer the benefits of scalability and flexibility \cite{xie2015precise}. 

Note that the non-contact monitoring methods 2-4 all share one common theme---wireless/RF sensing. The success of such systems is based upon the fact that the human body has more than 60\% of water, which could efficiently reflect the wireless signals impinging on the body, and thus it enables the detection of various kinds of human activities. In other words, the signals reflected off the body once received convey information about body movement (as they vary proportionally with body movement). More precisely, different body movements lead to different (mostly unique)  received signal patterns, which are utilized by the ML and DL methods that extract the relevant features to do vitals estimation (e.g., heart rate, blood oxygen saturation level, breathing rate, etc.) \cite{xie2015precise}.  

One key benefit of RF sensing solutions is the enhanced coverage, and thus it alleviates many of the challenges faced by the camera-based solutions. Furthermore, RF-based remote sensing of vitals could be done beyond the walls, doors and windows (for sufficiently-low frequencies) \cite{wang2015understanding, yatani2012bodyscope}.

\subsection{Contributions} 

The key contributions of this work are as follows:

1) We propose hand-breathe, a novel non-contact method to detect breathing abnormalities. Specifically, the subject rests his/her hand on a table in between the transmit and receive horn antennas. The transmitted OFDM signal passes through the hand, and is collected at the receive end. The receiver extracts from the received signal the channel frequency response (CFR), i.e., fined-grained wireless channel state information (WCSI). The CFR/WCSI is then fed to various ML classifiers which eventually classify between different breathing abnormalities. Among all classifiers, linear support vector machine (SVM) classifier yielded the best accuracy which is 88.1\%.  

2) In addition to our proposed method (where a hand is exposed to RF signals), we also implement and test the state-of-the-art method \cite{benchmarkworkQammer} (where full chest is exposed to RF radiation). The performance comparison of the two methods reveals a trade-off, i.e., the accuracy of our proposed method is slightly inferior than the benchmark method, but our method results in minimal body exposure to RF radiation compared to the benchmark method.


\subsection{Outline} 
Section II summarizes the related work. Section III introduces the system model. Section IV outlines the essential details of the proposed hand-breathe method. Section V presents results of performance evaluation of the ML classifiers. Section VI concludes the paper. 

\section{Related Work}

This section provides a compact yet comprehensive review of the related work on contact-less RF sensing methods that monitor breathing abnormalities.

{\it 1) Camera-based respiration monitoring systems:}
Such systems utilize either thermal imaging or depth cameras to monitor the breathing abnormalities. They exploit the fact that during breathing the temperature around the nose changes, and use infrared thermography for sensing breathing abnormalities  \cite{mei2011robust}. Similarly, the depth cameras and video cameras are also used for capturing the breathing activities of humans. Such methods have some drawbacks, e.g. susceptibility to heat and high computational cost in the case of thermal and depth cameras respectively \cite{sato2006non}.

{\it 2) Radar-based respiration monitoring systems:}
Various kinds of radar modules, e.g., frequency-modulated continuous-wave (FMCW) radar, ultra-wideband (UWB) pulse radar, continuous-wave (CW) Doppler radar, ultrasonic-based radar, and mm-wave radar are among the RF-based sensing methods for detection of breathing abnormalities. For example, the FMCW radars work on the classical principle of sending radar signals with varying frequency over time, and the signals reflected-off the subject are recorded and processed. The reflected signal typically undergoes a change in amplitude and phase as a result of the subject's breathing activity. This helps the radar modules to detect breathing abnormalities by utilizing the range and Doppler principles \cite{van2016wireless}. As another example, UWB radar sends short-duration pluses toward the target subject and captures the weak reflected signal that is delayed in time. This signal is used to calculate the distance between the radar and the target, which in turn helps to monitor the breathing abnormalities \cite{ali2017one}. CW Doppler radar, on the other hand, transmits sinusoidal RF signals which get modulated by chest and abdomen movement, and are received subsequently. Finally, self-injection locked ultrasonic radar has also been used to measure breathing abnormalities \cite{park2007arctangent,yu2019highly}. Radar-based technologies provide a non-contact solution for the sensing of breathing abnormalities but they typically require high-cost equipment, and are radiation hazards.

{\it 3) Wi-Fi-based respiration monitoring systems:}
During the last decade, Wi-Fi has become one of the most extensively used technology in our daily lives. WiFi systems utilize the radio signal strength (RSS) and channel state information (CSI) of the received signals after they reflect-off the subject, in order to monitor the breathing of the subject. The multipath effect captured by the RSS provides the ML algorithms with a coarse feature which they capitalize upon to detect breathing abnormalities. On the other hand, CSI provides fine-grained information about the body movement, and captures power attenuation due of multipath \cite{kaltiokallio2014non,chen2017tr}. As mentioned earlier, the Wi-Fi-based solutions require a low initial capital, have a low operational cost, offer ubiquitous operations (due to standardized and easily available hardware), but these systems lack the flexibility (due to rigid design of the hardware).

{\it 4) SDR-based respiration monitoring systems:}
All the SDR-based methods for monitoring breathing abnormalities rely upon the observation that the (amplitude and phase) variations of the signals reflected-off the subject are directly proportional to the tiny chest movements during a breathing cycle (exhale and inhale) \cite{rehman2021rf,al2019wireless,rehman2021improving}.  
More precisely, the SDR-based methods utilize the WCSI which contain the combined effect of reflection, scattering, shadowing, scattering, and power decay with increasing distances, all due to the motion of body \cite{wang2021csi,khan2021tracking}. Many works have been reported which claimed to detect large-scale movements like standing, walking falling, and small-scale moments such as keystroke, gesture recognition, rapid eye moment (REM), and breathing \cite{wang2020csi,wang2020resilient,zeng2020multisense,liu2020wi,niu2021wimonitor}. Finally, the RF sensing systems based on SDRs provide various advantages, e.g., scalability, flexibility, and reliability.


\section{System Model}
\label{sec:sys-model}

\begin{figure}[ht]
\begin{center}
	\includegraphics[width=9cm,height=5cm]{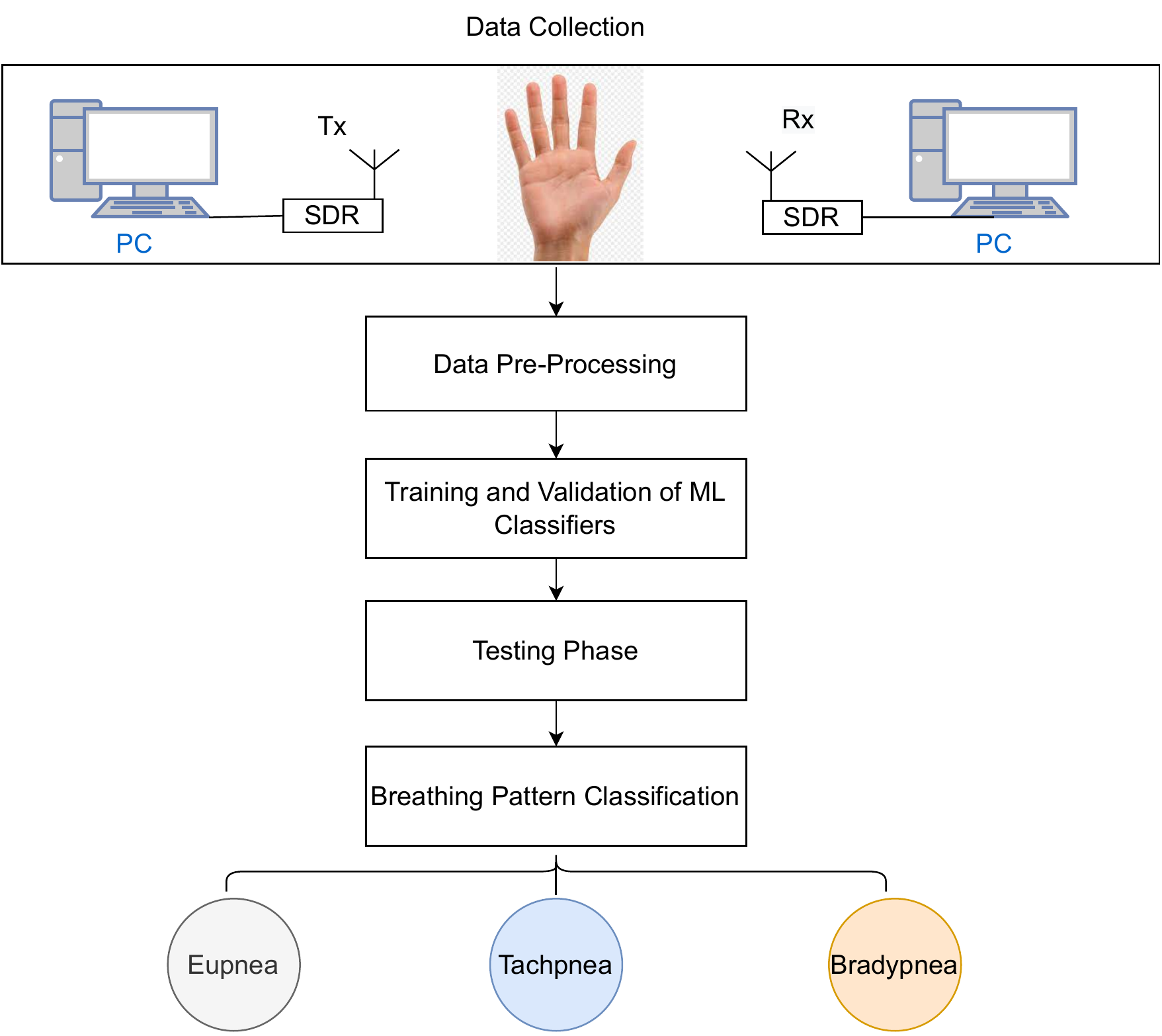} 
\caption{The system Model (flowchart) of the proposed SDR-based, ML-empowered non-contact method for breathing abnormalities detection.}
\label{fig:sysmodel}
\end{center}
\end{figure}

Fig. \ref{fig:sysmodel} shows the system model (flowchart) for our proposed SDR-based, ML-empowered system for detection of breathing abnormalities. Below, we provide some succinct details about the specific SDR platform  used, the Matlab-based software interface used to program the SDRs, the nature of experiments conducted, and Matlab classification learner app that we used for training and validation of various ML classifiers.

The USRP B210 SDR:
The proposed method utilizes a hardware platform that is comprised of two PCs, two USRP SDRs, and two directional horn antennas (see Fig. \ref{fig:sysmodel}). In particular, we chose USRP B210 module by National Instruments due to its wide frequency range of operation and affordable cost. The USRP B210 comes with a Spartan-6 field-programmable gate array, integrated RF daughterboards, and could simultaneously support two transmit and two receive data streams.

Software interface for the USRP SDR:
The data collection for training of the ML classifiers was done using the platform shown in Fig. \ref{fig:sysmodel}. For this purpose, Matlab communication system toolbox was used that contains an add-on that enables the USRP SDR to exchange real-time data with the host PC. Additionally, the whole flowgraphs for the transmitter and receiver were made in Simulink by utilizing various building blocks (e.g., modulators, demodulators, etc.) from communication system toolbox. 

The experiment:
OFDM symbols (with QAM modulation on each sub-carrier) synthesized in Matlab at the host PC were sent to the USRP B210 through an Ethernet cable using MATLAB SDRu block. The transmitted signal passed through the hand of subject placed in between the transmit and receive antennas, and was subsequently received at the receiver end. The receiver then extracted the WCSI from the received signal. This raw data was later used to train the ML classifiers.

Classification learner application of Matlab:
The classification learner application of Matlab was used to train various classification models. It has various options for analyzing the data, feature selection, selection of a validation scheme, training and evaluation of ML models.

\section{The proposed method}
\label{sec:method}

This section outlines the essential details of the proposed method that utilizes a pair of USRP SDRs to classify the breathing abnormalities with the aid of various ML algorithms in a contact-less fashion. Specifically, we describe in detail the following core steps of the proposed method: the classification problem at hand, the design of the USRP transmitter and receiver in baseband, the details of the experiments done, the data acquisition procedure, pre-processing of the data, and finally the training and testing of the ML algorithms. 

\subsection{The breathing pattern classification problem}
This work reformulates the breathing abnormalities detection problem into a classification problem with three classes: normal breathing, fast breathing and slow breathing. Some pertinent details for each breathing class are as follows.

Eupnea: Normal breathing activity is known as Eupnea. The normal breathing rate remains in the range of 12 to 20 breaths per minute. Traditionally, a healthy lifestyle and balanced diet have been recommended for normal breathing.

Tachypnea: Tachypnea occurs when an individual breathes faster than normal. It may be caused by anxiety, shock, exercise, and symptoms of lung disease.

Bradypnea: Slower breathing as compared to normal breathing is known as Bradypnea. The onset of Bradypnea typically indicates that the body is not getting enough Oxygen. It could also poin to other problems such as Carbon monoxide poisoning, head injury, metabolic disorder, and sleep apnea.

\subsection{Design of USRP transmitter and receiver in baseband}

Both transmit and receive flowgraphs are made in Simulink. Additionally, the communication system toolbox of Matlab is used in order to interface the USRP SDR to the host PC. 

\begin{figure*}
\hfill
\subfigure[The transmitter flowgraph]{\includegraphics[scale=0.18]{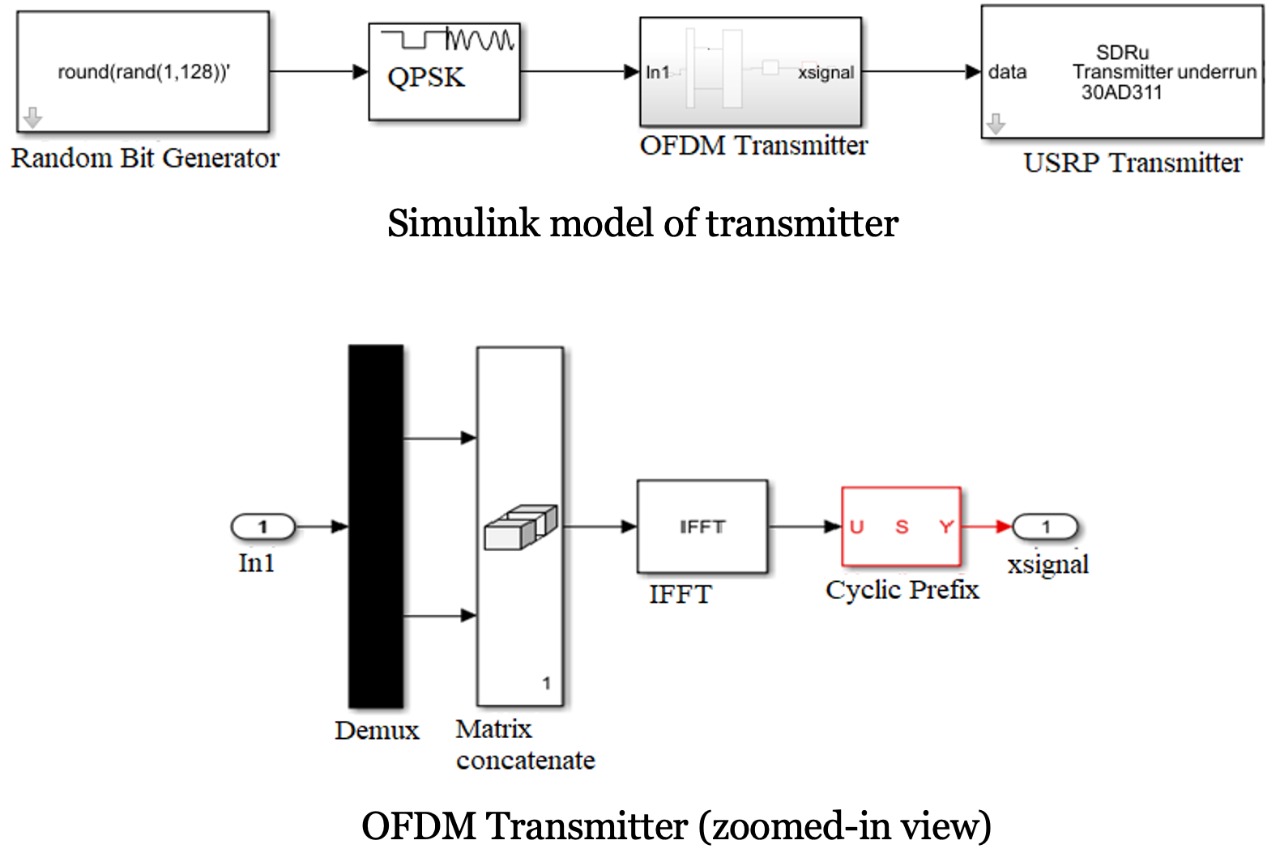}}
\hfill
\subfigure[The receiver flowgraph]{\includegraphics[scale=0.18]{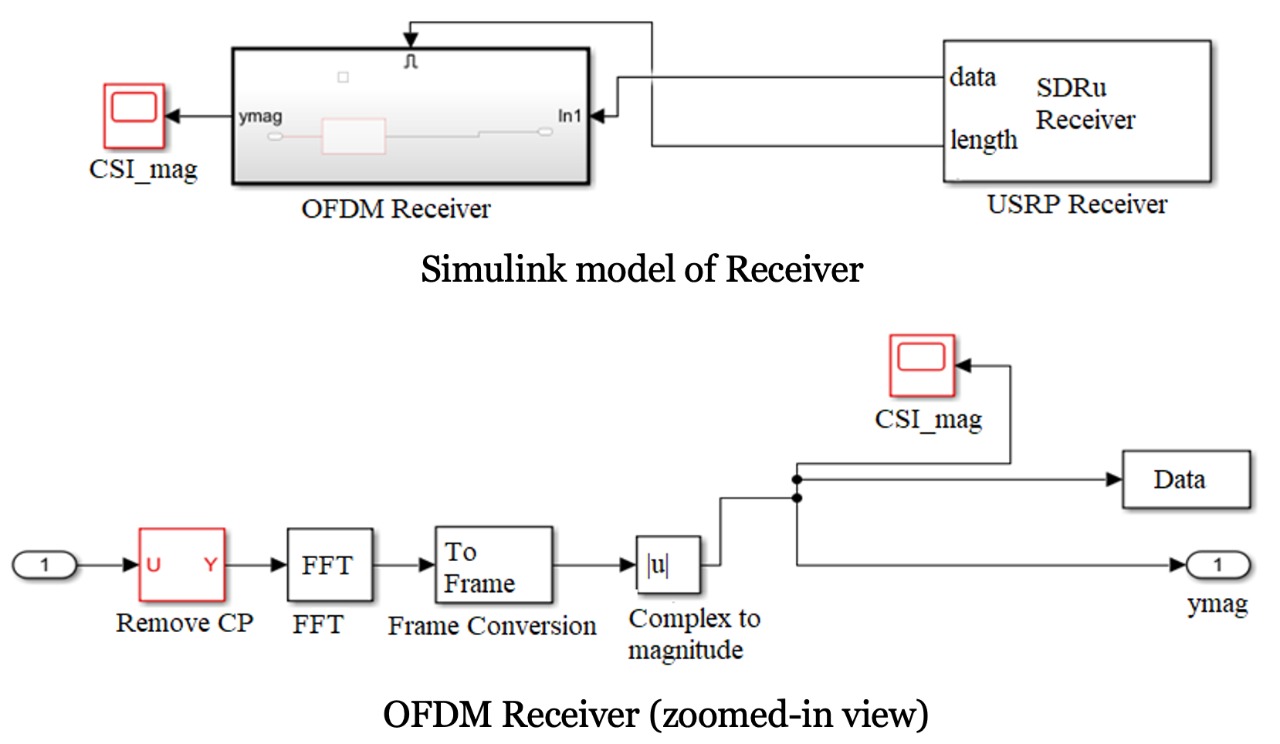}}
\hfill
\caption{The Simulink flowgraphs for USRP transmitter and USRP receiver}
\label{fig:flowg}
\end{figure*}

Transmitter flowgraph: Random bits (of chunk size 128 bits) are generated by means of the random bit generator block, for each OFDM frame. The bits are converted into symbols by the QPSK modulator block. Then, the serial data is converted into parallel and 64-point IFFT of the input symbols is taken. Subsequently, the last 16 samples of the current OFDM symbol are appended in the beginning as cyclic prefix (CP), which makes each OFDM symbol 80 samples long. The gain of transmit horn antenna is set to 40 dB. The SDRu sink block is employed to configure the hardware parameters of the transmit USRP SDR (see Table \ref{table:1}), and to transfer baseband samples to USRP B210. Fig. \ref{fig:flowg} (a) shows the transmitter flowgraph used to program the transmit USRP SDR.

Receiver flowgraph: The SDRu source block is employed to configure the hardware parameters of the receive USRP SDR (see Table \ref{table:1}), and to receive baseband samples from USRP B210. The gain of receive horn antenna is set to 40 dB. The first step of the receive flowgraph is to identify the start of the OFDM frame in order to remove the 16 samples corresponding to CP. This is followed by an FFT block that returns the noisy OFDM symbol. Next, complex-valued CFR is extracted, which is one of the many manifestations of the WCSI. The variation in WCSI/CFR arises due to channel effects, and is investigated to sense the breathing abnormalities. Fig. \ref{fig:flowg} (b) shows the transmitter flowgraph used to program the receive USRP SDR.

Table \ref{table:1} lists all the important USRP configuration parameters along with their values.

\begin{table}[h!]
\centering
\begin{tabular}{|c| c|} 
 \hline
 Parameter & Type/Value
	
\\
\hline

Bits per OFDM frame &	$128$ \\
Bits per symbol &	$2$ \\
Coding scheme &	Gray coding
 \\
Modulation scheme &	QPSK \\
No. of OFDM subcarriers &	$64$\\
Data subcarriers &	$52$\\
Pilot subcarriers &	$12$\\
Size of FFT/IFFT &	$64$ points\\
Size of cyclic prefix	& $16$\\
Sampling rate&	$1000$ samples/sec\\
Antenna type&	directional horn \\
USRP B210 frequency range&	$70$ MHz - $6$ GHz\\
Centre frequency&	$5.23$ GHz\\
Clock source \& PPS source &	Internal\\
Internal clock rate &	$200$ MHz\\
Interpolation factor (at Tx) &	$250$\\
Decimation factor (at Rx) &	$250$\\
Transmitter gain (at Tx and Rx) &	$40$ dB\\

 \hline
\end{tabular}
\caption{Configuration of transmit USRP and receive USRP: important parameters}
\label{table:1}
\end{table}

\subsection{Experimental Setup}
\begin{figure}[ht]
\begin{center}
	\includegraphics[width=8cm,height=4cm]{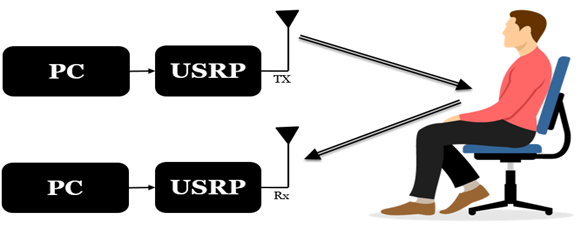} 
\caption{The experimental setup of the benchmark method that aims to capture the small-scale chest movements for breathing abnormality detection.}
\label{fig:method1}
\end{center}
\end{figure}

\begin{figure}[ht]
\begin{center}
	\includegraphics[width=9cm,height=3cm]{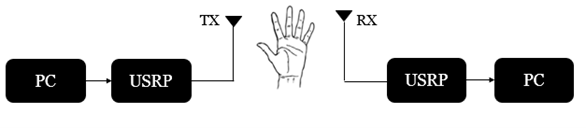} 
\caption{The experimental setup of proposed hand-breathe method that aims to capture the minute rhythmic movements of the hand for breathing abnormality detection.}
\label{fig:methodour}
\end{center}
\end{figure}

Once the design of transmitter and receiver flowgraphs was finalized, we proceeded to perform real-time experiments on volunteer subjects in order to do data acquisition for training of various ML classifiers later on. To be comprehensive in our analysis, we performed two sets of experiments. During first set of experiments, we collected data according to the protocol described in the state-of-the-art method \cite{benchmarkworkQammer} whereby the transmitter SDR impinges the signal onto the chest of the subject, while the receiver SDR collects the signal reflected-off the body (see Fig. \ref{fig:method1}). The second set of experiments represents the proposed method where the subject places his/her hand on a table such that the transmitted signal passed through the hand, and the weak signal that penetrates through the hand is collected at the receive end (see Fig. \ref{fig:methodour}). 

It is worth mentioning that first set of experiments was done in order to utilize the work \cite{benchmarkworkQammer} as a benchmark to assess the performance of our proposed method against it. 

{\it The benchmark method (chest movement-based experiment):} 
Two USRP radios each connected to a PC are placed on a table with their directional horn antennas pointing towards the subject who sits on a chair at about a distance of 80 cm from the table, as shown in Fig \ref{fig:method1}. The transmit horn antenna impinges the OFDM signal on the chest of the subject, while the receive horn antenna collects the signal reflected-off the chest of the subject. With this, the aim is to capture the small-scale movement of the chest (due to respiration activity). 

{\it The proposed method (hand movement-based experiment):} 
Two USRP radios each connected to a PC are placed on a table with their directional horn antennas (separated by 60 cm) pointing towards each other, as shown in Fig \ref{fig:methodour}. The subject sits on a chair nearby the table and places his/her hand on the table in between the transmit and receive horn antennas. As before, the transmit horn antenna impinges the OFDM signal on the hand of the subject, while the receive horn antenna collects the weak signal that successfully penetrates through the hand of the subject. With this, we aim is to capture the faint movements of the hand (due to respiration activity). Specifically, the proposed method intends to capitalize on the fact the hand being connected to the shoulder undergoes a minute periodic movement due to movement of the chest.  


\subsection{Data Acquisition}

\begin{figure*}
\hfill
\subfigure[Data collection for benchmark method]{\includegraphics[scale=0.2]{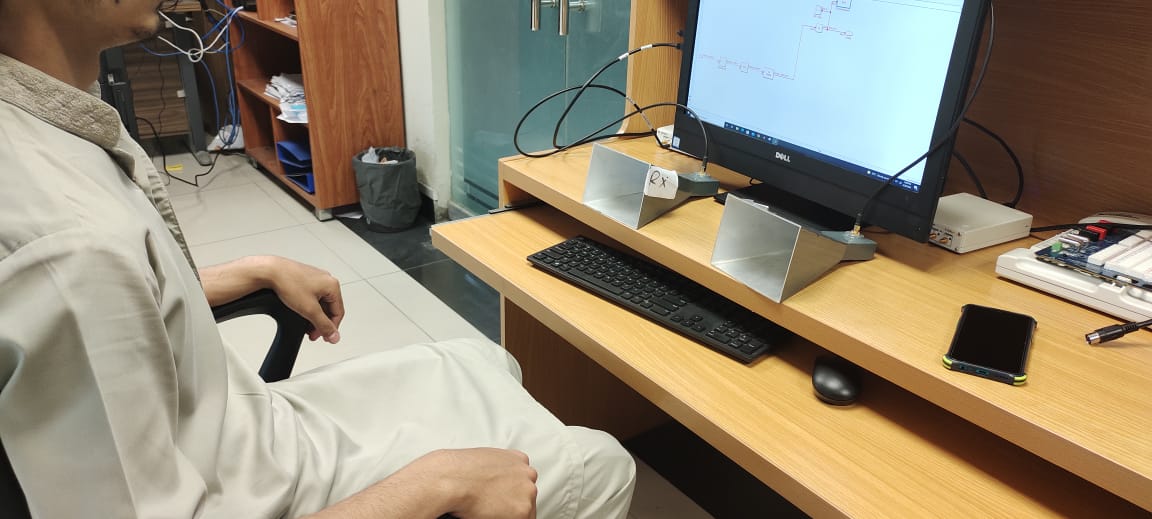}}
\hfill
\subfigure[Data collection for proposed method]{\includegraphics[scale=0.2]{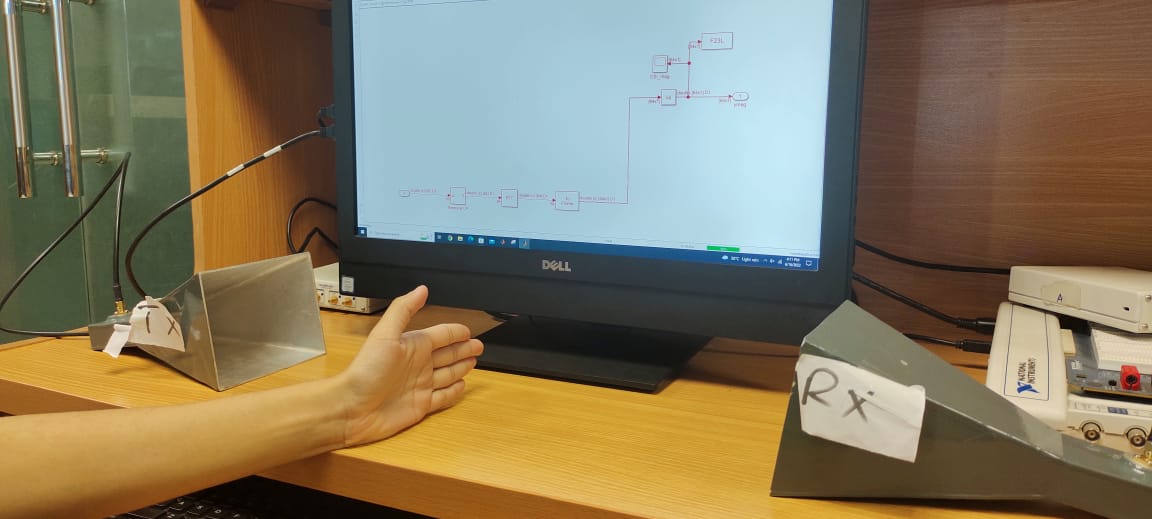}}
\hfill
\caption{Data acquisition campaign: benchmark method (left), and proposed method (right)}
\label{fig:real_exp_pics}
\end{figure*}

{\it Data acquisition for benchmark method (chest movement-based): } 
Data was collected in the lab environment. Four healthy volunteers (all males) participated in the data collection campaign. Each subject was requested to perform three different breathing activities (Eupnoea, Tachypnea, and Bradypnea) artificially. In other words, each subject breathed normally for Eupnoea experiment, underwent fast breathing for Tachypnea experiment, and performed slow breathing for Bradypnea experiment. Furthermore, the experiment for each breathing activity for each subject was repeated five times. Each experiment (transmission) lasted for 30 seconds. This led to a total of $4\times 3\times 5=60$ experiments. The data collected after each experiment consisted of the complex-valued CFR (a manifesttaion of the WSCI). Table \ref{table:datasetref} summarizes the key statistics of the data collection done for the chest movement-based experiment. Fig. \ref{fig:real_exp_pics} (a) shows the experimental setup used for data acquisition campaign for the benchmark method.

{\it Data acquisition for proposed method (hand movement-based): }
The data acquisition campaign for hand movement-based experiment is completely identical to the data collection campaign for the first experiment. That is, 4 volunteers performed 3 breathing activities while repeating each activity 5 times, thus giving rise to a total of 60 experiments. However, note that the subjects were advised to keep their hand (under test) static on the table between the transmit and receive horn antenna, while performing different breathing activities. This was to make sure that there are no motion-induced artefacts in the data being collected (which would in turn help in capturing the very subtle/minute movement of hand due to respiration). Table \ref{table:datasetref} could again be referred to, to get a quick glimpse at the key statistics of the data collected for the hand-movement based experiment. Fig. \ref{fig:real_exp_pics} (b) shows the experimental setup used for data acquisition campaign for the proposed method.

\begin{table}[h!]
\centering
\begin{tabular}{|c| c|} 
 \hline
 Parameter & Value
	
\\
\hline
Number of subjects&	$4$ \\
Number of breathing activities performed &	$3$ \\
Number of times each activity was performed &	$5$ \\
Duration of each experiment &	$30$ sec \\
Total number of experiments&	$60$ \\
Number of USRP radios used &	$2$ \\
Number of PCs used &	$2$ \\

 \hline
\end{tabular}
\caption{Pertinent details of data collection campaign for both methods (the proposed method and the benchmark method) }
\label{table:datasetref}
\end{table}

\subsection{Data Pre-processing \& Training of Machine Learning Classifiers }  

{\it Data Preprocessing:}
First of all, the raw data collected from each experiment was manually cleaned by zero-padding of the missing data, and discarding the the corrupted (highly noisy) data. Next, the data was manually annotated/labeled. Finally, some further reshaping/resizing of the data was done for each breathing experiment. Eventually the data corresponding to all experiments for each of the two methods (the proposed method and the benchmark method) were concatenated to construct a single file representing the labeled data. 

{\it Training \& validation of ML classifiers: }
Three supervised machine learning classifiers, i.e., k nearest neighbors (k-NN), support vector machine (SVM), and decision tree (DT) were trained on the two labeled datasets (corresponding to the proposed method and the benchmark method). For validation, the K-fold cross-validation approach was implemented in order to avoid the over-fitting problem. Recall that in the cross-validation approach, the dataset is randomly divided into smaller chunks for testing and training. For example, if K is 5, then 80$\%$ of the data is used for training, while remaining 20$\%$ of the data is used for testing/validation purpose. Further, with K=5, the classifier model will train 5 times.

\section{Performance Evaluation}

This section discusses the performance of the proposed SDR-based, ML-empowered non-contact method that captures the subtle but periodic movements of the hand for breathing abnormality detection. The accuracy of each classifier is evaluated as follows:
\begin{align}
   & \mathrm{Accuracy}=\frac{\mathrm{Correct \; prediction}}{\mathrm{Total \; observations}}\times 100
\\
&\mathrm{Accuracy}=\frac{T_n+T_p}{T_n+T_p+F_n+F_p} \times 100  
\end{align}
where $T_n$ is true negative, $T_p$ is true positive, $F_n$ is false negative and $F_p$ is false positive, respectively. Additionally, we also construct the confusion matrices in order to investigate the test accuracy of the three ML classifiers. 

\subsection{Performance of Benchmark method (chest movement-based)}
Recall that the benchmark method relies upon the small-scale chest movements due to respiration for breathing pattern classification.

We first discuss the performance of the k-NN classifier. Since k represents the number of neighbors the algorithm computes the distance of the new data point from, we studied the performance of k-NN for three different values of k. Accordingly, we obtained an overall accuracy of 80$\%$ for fine k-NN (k=1), 67.8$\%$ for medium k-NN (k=9), and 75.6$\%$ (k=3). The detailed confusion matrix showing the performance breakup of the k-NN for the three breathing patterns and for three different values of k is shown in Table \ref{table:knnref}. 

We also tested the SVM classifier (with both linear and cubic kernel) and obtained an overall accuracy of 93.6 $\%$ for linear SVM, and 95.9$\%$ for cubic SVM. Furthermore, the detailed confusion matrix showing the performance breakup of the linear SVM and cubic SVM for the three breathing patterns is presented in Table \ref{table:svmdtref}. Last but not the least, the DT classifier yields the lowest overall accuracy of 51.7$\%$. 

Table \ref{table:refmoverall} provides a crisp performance comparison of all the three classifiers. The SVM classifier (with cubic kernel) performs the best with an accuracy of $95.9$\%, for the benchmark method (that relies upon the small-scale chest movements due to respiration for breathing pattern classification).

\begin{table}[h!]
\centering
\begin{tabular}{|c| c| c| c|c|} 
 \hline
 Algorithm &
	Actual /Predicted&	Eupnea &	Tachypnea	&Bradypnea
\\
\hline

 &	Eupnea &	\cellcolor{blue!25}82.66 & 5.70 &	11.64 \\
k-NN
(k=1) &		Tachypnea &	9.22 &	\cellcolor{blue!25}76.41 &	14.38 \\

 &	Bradypnea &	10.78 &	8.28 &	\cellcolor{blue!25}80.94 \\
\hline 

 &	Eupnea &	\cellcolor{blue!25}75.16 &	7.89 &	16.95 \\
k-NN
(k=9) &	Tachypnea&	18.44 &	\cellcolor{blue!25}60.16 &	21.41 \\

 &	Bradypnea &	21.80 &	10.08 &	\cellcolor{blue!25}68.13 \\
\hline 

	& Eupnea &	\cellcolor{blue!25}82.66 &	4.69 &	12.66 \\
k-NN
(k=3) &	Tachypnea &	14.69 &	\cellcolor{blue!25}68.75 &	16.56 \\

 &	Bradypnea &	17.58 &	7.66 &	\cellcolor{blue!25}74.77 \\
 \hline
\end{tabular}
\caption{Confusion matrix of k-NN algorithm for the benchmark method (chest movement-based). Total number of observations for each breathing class is 1280. }
\label{table:knnref}
\end{table}

\begin{table}[h!]
\centering
\begin{tabular}{|c| c| c| c|c|} 
 \hline
 Algorithm &
	Actual /Predicted&	Eupnea &	Tachypnea	& Bradypnea
\\
\hline
 &	Eupnea &	\cellcolor{blue!25}90.0 & 1.02 &	8.98 \\
Linear SVM &		Tachypnea &	1.17 &	\cellcolor{blue!25}92.5 &	6.33 \\
 &	Bradypnea &	0.63 &	1.02 &	\cellcolor{blue!25}98.36 \\
\hline 
 &	Eupnea &	\cellcolor{blue!25}96.09 &	0.39 &	3.52 \\
Cubic SVM&	Tachypnea &	1.72 &	\cellcolor{blue!25}93.52 &	4.77 \\
&	Bradypnea &	0.70 &	1.09 &	\cellcolor{blue!25}98.2 \\
\hline 
	& Eupnea &	\cellcolor{blue!25}48.05 &	5.94 &	46.02 \\
DT &	Tachypnea &	13.52 &	\cellcolor{blue!25}32.11 &	54.38 \\
&	Bradypnea &	17.73 &	7.19 &	\cellcolor{blue!25}75.08 \\
 \hline
\end{tabular}
\caption{Confusion matrix of SVM and DT classifiers for the benchmark method (chest movement-based). Total number of observations for each breathing class is 1280.}
\label{table:svmdtref}
\end{table}

\begin{table}[h!]
\centering
\begin{tabular}{|c| c|} 
 \hline
 Algorithm &
	Accuracy
\\
\hline
Fine k-NN (k=1) & 80$\%$ \\
Medium k-NN (k=9) & 67.8$\%$ \\
Medium k-NN (k=3) & 75.4$\%$ \\
\hline 
Linear SVM & 93.6$\%$ \\
Cubic SVM & 95.9$\%$ \\
\hline 
DT & 51.7$\%$ \\
 \hline
\end{tabular}
\caption{Performance comparison of all the three classifiers for the benchmark method (chest movement-based).}
\label{table:refmoverall}
\end{table}

\subsection{Performance of Proposed method (hand movement-based)}
Recall that the proposed method capitalizes on very minute but rhythmic movements of the hand due to respiration for breathing pattern classification.

Table \ref{table:confmatprop} provides the detailed confusion matrix showing the performance breakup of the fine k-NN (with k=1), medium k-NN (with k=3), linear SVM, quadratic SVM and DT classifiers. We note that both the fine k-NN (k=1) and medium k-NN (k=3) perform poorly as they obtain an overall accuracy of 45.7$\%$ and 45.9\%, respectively. On the other hand, linear SVM and quadratic SVM perform very well with an overall accuracy of 88.1 $\%$ and 85.7$\%$, respectively. The DT algorithm performs the worst with an overall accuracy of 40.8$\%$. 

Table \ref{table:propmoverall} provides a compact summary of overall performance of the three classifiers. The SVM classifier (with linear kernal) performs the best with an overall accuracy of $88.1$\%, for the proposed method (which capitalizes on very minute but rhythmic movements of the hand due to respiration for breathing pattern classification).

\begin{table}[h!]
\centering
\begin{tabular}{|c| c| c| c|c|} 
 \hline
 Algorithms &
	Actual /Predicted&	Eupnea &	Tachypnea	&Bradypnea
\\
\hline
&	Eupnea &	\cellcolor{blue!25}46& 23&	30 \\
k-NN (k=1)&		Tachypnea&	29&	\cellcolor{blue!25}44&	27 \\
&	Bradypnea&	29&	24&	\cellcolor{blue!25}48 \\
\hline
&	Eupnea &	\cellcolor{blue!25}44& 23&	33 \\
k-NN (k=3)&		Tachypnea&	28&	\cellcolor{blue!25}44&	28 \\
&	Bradypnea&	29&	22&	\cellcolor{blue!25}49 \\
\hline
&	Eupnea &	\cellcolor{blue!25}94.60& 0.62&	4.8 \\
Linear SVM &		Tachypnea&	8.35&\cellcolor{blue!25}	80.31&	11.32 \\
&	Bradypnea&	8.98&	1.71&	\cellcolor{blue!25}89.30 \\
\hline 
&	Eupnea&	\cellcolor{blue!25}87.18&	4.76&	8.04 \\
Quad. SVM&	Tachypnea&	6.40&	\cellcolor{blue!25}84.30&	9.30 \\
&	Bradypnea&	9.21&	5.15&	\cellcolor{blue!25}85.62 \\
\hline 
	&Eupnea&	\cellcolor{blue!25}91.32&	3.30&	5.40 \\
DT &	Tachypnea&	80.40&	\cellcolor{blue!25}13.12&	6.48\\
&	Bradypnea&	76.87&	5.08&	\cellcolor{blue!25}18.04 \\
 \hline
\end{tabular}
\caption{Confusion matrix of the three classifiers for the proposed method (hand movement-based). Total number of observations for each breathing class is 1280.}
\label{table:confmatprop}
\end{table}
\begin{table}[h!]
\centering
\begin{tabular}{|c| c|} 
 \hline
 Algorithm &
	Accuracy
\\
\hline
Fine k-NN (k=1) & 45.9$\%$ \\
\hline 
Fine k-NN (k=3) & 45.7$\%$ \\
\hline 
Linear SVM & 88.1$\%$ \\
Quadratic SVM & 85.7$\%$ \\
\hline 
DT & 40.8$\%$ \\
 \hline
\end{tabular}
\caption{Performance comparison of all the three classifiers for the proposed method (hand movement-based).}
\label{table:propmoverall}
\end{table}

\subsection{Discussions}
We presented in detail the performance results of the three classifiers for both methods (i.e., the benchmark method that exploits the chest movement, and our proposed method that capitalizes on subtle but rhythmic hand movements). The results point towards a tradeoff: our proposed method has slightly lower accuracy than the benchmark method (95.9\% vs. 88.1\%), but our method will expose the subjects to least amount of potentially harmful radiation (full chest vs. a hand only). We further notice that the performance of proposed method may further be increased by implementing the state-of-the-art deep learning methods which will require a very large dataset in order to train (leading to a very extensive data collection campaign), but do have the potential to supersede the performance of the benchmark method.  

A key benefit of the proposed method is that it could lead to the development of a smart mobile health (m-health) solution that could be deployed in remote areas far away from mega cities in order to ensure the comprehensive health monitoring of the people in those areas. Additionally, the proposed method could prove to be very useful for rapid and non-contact testing of masses for their respiratory performance during the outbreak of a pandemic like covid19.

\section{Conclusion \& Future Work}
\label{sec:conclusion}

This work proposed hand-breathe, a novel method for non-contact monitoring of breathing abnormalities. Specifically, we utilized a pair of USRP radios which exchanged known OFDM symbols while the subject placed his/her hand between the transmit and receive antennas. Subsequently, the receiver extracted the CFR/WCSI that captured the small-scale hand movements of the subject due to respiration. Eventually, three ML algorithms were used to classify the breathing performance of the subject into three different classes: normal, fast, and slow. Maximum overall accuracy of 88.1\% was achieved with the Linear SVM ML algorithm. Furthermore, the current state-of-the-art method (chest movements-based method) was also implemented as a benchmark method, and its performance was compared against the proposed method. This led us to discver a trade-off: our proposed method has slightly lower accuracy than the benchmark method (95.9\% vs. 88.1\%), but our method exposes the subjects to least amount of potentially harmful radiation (full chest vs. a hand only).

This work opens up many interesting directions for the future work. 
For example, another dataset could be collected from actual patients suffering from various respiratory diseases, in order to increase the generalization capability of the proposed ML algorithms. The testing and exploration of the coverage limits of the proposed method (by increasing the distance between the transceiver and subjects) is another interesting problem to look at. One may also consider extending the proposed method to monitor the breathing performance of multiple subjects simultaneously. Finally, the performance of proposed hand-breathe method may further be increased by implementing the state-of-the-art deep learning methods which will require a very large dataset in order to train (leading to a very extensive data collection campaign), but could supersede the performance of the proposed hand-breathe method.





\footnotesize{
\bibliographystyle{IEEEtran}
\bibliography{references}

\begin{thebibliography}{10}
\providecommand{\url}[1]{#1}
\csname url@rmstyle\endcsname
\providecommand{\newblock}{\relax}
\providecommand{\bibinfo}[2]{#2}
\providecommand\BIBentrySTDinterwordspacing{\spaceskip=0pt\relax}
\providecommand\BIBentryALTinterwordstretchfactor{4}
\providecommand\BIBentryALTinterwordspacing{\spaceskip=\fontdimen2\font plus
\BIBentryALTinterwordstretchfactor\fontdimen3\font minus
  \fontdimen4\font\relax}
\providecommand\BIBforeignlanguage[2]{{%
\expandafter\ifx\csname l@#1\endcsname\relax
\typeout{** WARNING: IEEEtran.bst: No hyphenation pattern has been}%
\typeout{** loaded for the language `#1'. Using the pattern for}%
\typeout{** the default language instead.}%
\else
\language=\csname l@#1\endcsname
\fi
#2}}

\bibitem{goldhill2005physiologically}
D.~Goldhill, A.~McNarry, G.~Mandersloot, and A.~McGinley, ``A
  physiologically-based early warning score for ward patients: the association
  between score and outcome,'' \emph{Anaesthesia}, vol.~60, no.~6, pp.
  547--553, 2005.

\bibitem{levine2022global}
S.~M. Levine and D.~D. Marciniuk, ``Global impact of respiratory disease: What
  can we do, together, to make a difference?'' \emph{Chest}, vol. 161, no.~5,
  pp. 1153--1154, 2022.

\bibitem{subbe2003effect}
C.~Subbe, R.~Davies, E.~Williams, P.~Rutherford, and L.~Gemmell, ``Effect of
  introducing the modified early warning score on clinical outcomes,
  cardio-pulmonary arrests and intensive care utilisation in acute medical
  admissions,'' \emph{Anaesthesia}, vol.~58, no.~8, pp. 797--802, 2003.

\bibitem{fieselmann1993respiratory}
J.~F. Fieselmann, M.~S. Hendryx, C.~M. Helms, and D.~S. Wakefield,
  ``Respiratory rate predicts cardiopulmonary arrest for internal medicine
  inpatients,'' \emph{Journal of general internal medicine}, vol.~8, no.~7, pp.
  354--360, 1993.

\bibitem{smith1993fuel}
K.~R. Smith, ``Fuel combustion, air pollution exposure, and health: the
  situation in developing countries,'' \emph{Annual Review of Energy and the
  Environment}, vol.~18, no.~1, pp. 529--566, 1993.

\bibitem{farida2020quality}
F.~Farida, A.~Retnosari, and R.~Maksum, ``Quality of antibiotic prescribing for
  respiratory tract disease in primary healthcare centers in tegal regency,
  central java, indonesia,'' \emph{Indonesian Journal of Clinical Pharmacy},
  vol.~9, no.~2, pp. 95--104, 2020.

\bibitem{tu2012computational}
J.~Tu, K.~Inthavong, and G.~Ahmadi, \emph{Computational fluid and particle
  dynamics in the human respiratory system}.\hskip 1em plus 0.5em minus
  0.4em\relax Springer Science \& Business Media, 2012.

\bibitem{garcia2018computational}
M.~Garcia-Gasulla, M.~Josep-Fabrego, B.~Eguzkitza, and F.~Mantovani,
  ``Computational fluid and particle dynamics simulations for respiratory
  system: Runtime optimization on an arm cluster,'' in \emph{Proceedings of the
  47th International Conference on Parallel Processing Companion}, 2018, pp.
  1--8.

\bibitem{zhang2016respiratory}
Z.~Zhang, ``Respiratory laryngeal coordination in airflow conservation and
  reduction of respiratory effort of phonation,'' \emph{Journal of Voice},
  vol.~30, no.~6, pp. 760--e7, 2016.

\bibitem{gramming1988relationship}
P.~Gramming, J.~Sundberg, S.~Ternstr{\"o}m, R.~Leanderson, and W.~H. Perkins,
  ``Relationship between changes in voice pitch and loudness,'' \emph{Journal
  of voice}, vol.~2, no.~2, pp. 118--126, 1988.

\bibitem{world2020clinical}
W.~H. Organization \emph{et~al.}, ``Clinical management of severe acute
  respiratory infection (sari) when covid-19 disease is suspected: interim
  guidance, 13 march 2020,'' World Health Organization, Tech. Rep., 2020.

\bibitem{saeed2021wireless}
U.~Saeed, S.~Y. Shah, A.~Zahid, J.~Ahmad, M.~A. Imran, Q.~H. Abbasi, and S.~A.
  Shah, ``Wireless channel modelling for identifying six types of respiratory
  patterns with sdr sensing and deep multilayer perceptron,'' \emph{IEEE
  Sensors Journal}, vol.~21, no.~18, pp. 20\,833--20\,840, 2021.

\bibitem{yatani2012bodyscope}
K.~Yatani and K.~N. Truong, ``Bodyscope: a wearable acoustic sensor for
  activity recognition,'' in \emph{Proceedings of the 2012 ACM Conference on
  Ubiquitous Computing}, 2012, pp. 341--350.

\bibitem{ertin2011autosense}
E.~Ertin, N.~Stohs, S.~Kumar, A.~Raij, M.~Al'Absi, and S.~Shah, ``Autosense:
  unobtrusively wearable sensor suite for inferring the onset, causality, and
  consequences of stress in the field,'' in \emph{Proceedings of the 9th ACM
  conference on embedded networked sensor systems}, 2011, pp. 274--287.

\bibitem{abdelnasser2015wigest}
H.~Abdelnasser, M.~Youssef, and K.~A. Harras, ``Wigest: A ubiquitous wifi-based
  gesture recognition system,'' in \emph{2015 IEEE conference on computer
  communications (INFOCOM)}.\hskip 1em plus 0.5em minus 0.4em\relax IEEE, 2015,
  pp. 1472--1480.

\bibitem{wang2014we}
G.~Wang, Y.~Zou, Z.~Zhou, K.~Wu, and L.~M. Ni, ``We can hear you with wi-fi!''
  in \emph{Proceedings of the 20th annual international conference on Mobile
  computing and networking}, 2014, pp. 593--604.

\bibitem{wang2016lifs}
J.~Wang, H.~Jiang, J.~Xiong, K.~Jamieson, X.~Chen, D.~Fang, and B.~Xie, ``Lifs:
  Low human-effort, device-free localization with fine-grained subcarrier
  information,'' in \emph{Proceedings of the 22nd Annual International
  Conference on Mobile Computing and Networking}, 2016, pp. 243--256.

\bibitem{lien2016soli}
J.~Lien, N.~Gillian, M.~E. Karagozler, P.~Amihood, C.~Schwesig, E.~Olson,
  H.~Raja, and I.~Poupyrev, ``Soli: Ubiquitous gesture sensing with millimeter
  wave radar,'' \emph{ACM Transactions on Graphics (TOG)}, vol.~35, no.~4, pp.
  1--19, 2016.

\bibitem{10.1145/2829988.2787487}
M.~Kotaru, K.~Joshi, D.~Bharadia, and S.~Katti, ``Spotfi: Decimeter level
  localization using wifi,'' \emph{SIGCOMM Comput. Commun. Rev.}, vol.~45,
  no.~4, pp. 269--282, aug 2015.

\bibitem{xie2015precise}
Y.~Xie, Z.~Li, and M.~Li, ``Precise power delay profiling with commodity
  wifi,'' in \emph{Proceedings of the 21st Annual international conference on
  Mobile Computing and Networking}, 2015, pp. 53--64.

\bibitem{wang2015understanding}
W.~Wang, A.~X. Liu, M.~Shahzad, K.~Ling, and S.~Lu, ``Understanding and
  modeling of wifi signal based human activity recognition,'' in
  \emph{Proceedings of the 21st annual international conference on mobile
  computing and networking}, 2015, pp. 65--76.

\bibitem{benchmarkworkQammer}
A.~M. Ashleibta, Q.~H. Abbasi, S.~A. Shah, M.~A. Khalid, N.~A. AbuAli, and
  M.~A. Imran, ``Non-invasive rf sensing for detecting breathing abnormalities
  using software defined radios,'' \emph{IEEE Sensors Journal}, vol.~21, no.~4,
  pp. 5111--5118, 2020.

\bibitem{mei2011robust}
X.~Mei and H.~Ling, ``Robust visual tracking and vehicle classification via
  sparse representation,'' \emph{IEEE transactions on pattern analysis and
  machine intelligence}, vol.~33, no.~11, pp. 2259--2272, 2011.

\bibitem{sato2006non}
I.~Sato and M.~Nakajima, ``Non-contact breath motion monitor ing system in full
  automation,'' in \emph{2005 IEEE Engineering in Medicine and Biology 27th
  Annual Conference}.\hskip 1em plus 0.5em minus 0.4em\relax IEEE, 2006, pp.
  3448--3451.

\bibitem{van2016wireless}
K.~Van~Loon, M.~Breteler, L.~Van~Wolfwinkel, A.~Rheineck~Leyssius, S.~Kossen,
  C.~Kalkman, B.~Van~Zaane, and L.~Peelen, ``Wireless non-invasive continuous
  respiratory monitoring with fmcw radar: a clinical validation study,''
  \emph{Journal of clinical monitoring and computing}, vol.~30, no.~6, pp.
  797--805, 2016.

\bibitem{ali2017one}
M.~Ali, H.~Shawkey, A.~Zekry, and M.~Sawan, ``One mbps 1 nj/b 3.5--4 ghz fully
  integrated fm-uwb transmitter for wban applications,'' \emph{IEEE
  Transactions on Circuits and Systems I: Regular Papers}, vol.~65, no.~6, pp.
  2005--2014, 2017.

\bibitem{park2007arctangent}
B.-K. Park, O.~Boric-Lubecke, and V.~M. Lubecke, ``Arctangent demodulation with
  dc offset compensation in quadrature doppler radar receiver systems,''
  \emph{IEEE transactions on Microwave theory and techniques}, vol.~55, no.~5,
  pp. 1073--1079, 2007.

\bibitem{yu2019highly}
S.-H. Yu and T.-S. Horng, ``Highly linear phase-canceling self-injection-locked
  ultrasonic radar for non-contact monitoring of respiration and heartbeat,''
  \emph{IEEE Transactions on Biomedical Circuits and Systems}, vol.~14, no.~1,
  pp. 75--90, 2019.

\bibitem{kaltiokallio2014non}
O.~Kaltiokallio, H.~Yi{\u{g}}itler, R.~J{\"a}ntti, and N.~Patwari,
  ``Non-invasive respiration rate monitoring using a single cots tx-rx pair,''
  in \emph{IPSN-14 Proceedings of the 13th International Symposium on
  Information Processing in Sensor Networks}.\hskip 1em plus 0.5em minus
  0.4em\relax IEEE, 2014, pp. 59--69.

\bibitem{chen2017tr}
C.~Chen, Y.~Han, Y.~Chen, H.-Q. Lai, F.~Zhang, B.~Wang, and K.~R. Liu,
  ``Tr-breath: Time-reversal breathing rate estimation and detection,''
  \emph{IEEE Transactions on Biomedical Engineering}, vol.~65, no.~3, pp.
  489--501, 2017.

\bibitem{rehman2021rf}
M.~Rehman, R.~A. Shah, M.~B. Khan, N.~A. AbuAli, S.~A. Shah, X.~Yang,
  A.~Alomainy, M.~A. Imran, and Q.~H. Abbasi, ``Rf sensing based breathing
  patterns detection leveraging usrp devices,'' \emph{Sensors}, vol.~21,
  no.~11, p. 3855, 2021.

\bibitem{al2019wireless}
A.~Al-Wahedi, M.~Al-Shams, M.~A. Albettar, S.~Alawsh, and A.~Muqaibel,
  ``Wireless monitoring of respiration and heart rates using
  software-defined-radio,'' in \emph{2019 16th International Multi-Conference
  on Systems, Signals \& Devices (SSD)}.\hskip 1em plus 0.5em minus 0.4em\relax
  IEEE, 2019, pp. 529--532.

\bibitem{rehman2021improving}
M.~Rehman, R.~A. Shah, M.~B. Khan, S.~A. Shah, N.~A. AbuAli, X.~Yang,
  A.~Alomainy, M.~A. Imran, and Q.~H. Abbasi, ``Improving machine learning
  classification accuracy for breathing abnormalities by enhancing dataset,''
  \emph{Sensors}, vol.~21, no.~20, p. 6750, 2021.

\bibitem{wang2021csi}
Z.~Wang, Z.~Huang, C.~Zhang, W.~Dou, Y.~Guo, and D.~Chen, ``Csi-based human
  sensing using model-based approaches: a survey,'' \emph{Journal of
  Computational Design and Engineering}, vol.~8, no.~2, pp. 510--523, 2021.

\bibitem{khan2021tracking}
M.~I. Khan, M.~A. Jan, Y.~Muhammad, D.-T. Do, C.~X. Mavromoustakis, E.~Pallis,
  \emph{et~al.}, ``Tracking vital signs of a patient using channel state
  information and machine learning for a smart healthcare system,''
  \emph{Neural Computing and Applications}, pp. 1--15, 2021.

\bibitem{wang2020csi}
X.~Wang, C.~Yang, and S.~Mao, ``On csi-based vital sign monitoring using
  commodity wifi,'' \emph{ACM Transactions on Computing for Healthcare},
  vol.~1, no.~3, pp. 1--27, 2020.

\bibitem{wang2020resilient}
------, ``Resilient respiration rate monitoring with realtime bimodal csi
  data,'' \emph{IEEE Sensors Journal}, vol.~20, no.~17, pp. 10\,187--10\,198,
  2020.

\bibitem{zeng2020multisense}
Y.~Zeng, D.~Wu, J.~Xiong, J.~Liu, Z.~Liu, and D.~Zhang, ``Multisense: Enabling
  multi-person respiration sensing with commodity wifi,'' \emph{Proceedings of
  the ACM on Interactive, Mobile, Wearable and Ubiquitous Technologies},
  vol.~4, no.~3, pp. 1--29, 2020.

\bibitem{liu2020wi}
W.~Liu, S.~Chang, Y.~Liu, and H.~Zhang, ``Wi-psg: Detecting rhythmic movement
  disorder using cots wifi,'' \emph{IEEE Internet of Things Journal}, vol.~8,
  no.~6, pp. 4681--4696, 2020.

\bibitem{niu2021wimonitor}
X.~Niu, S.~Li, Y.~Zhang, Z.~Liu, D.~Wu, R.~C. Shah, C.~Tanriover, H.~Lu, and
  D.~Zhang, ``Wimonitor: Continuous long-term human vitality monitoring using
  commodity wi-fi devices,'' \emph{Sensors}, vol.~21, no.~3, p. 751, 2021.

\end{thebibliography}
}

\vfill\break

\end{document}